\documentclass[conference]{IEEEconf}

\input epsf
\usepackage{cite}
\usepackage{graphicx}
\usepackage{multirow}
\usepackage{titlesec}
\usepackage{balance}
\usepackage{stfloats}
\usepackage{xcolor}
\usepackage{multirow}
\usepackage{hyperref}
\usepackage{tcolorbox}
\usepackage{listings}
\usepackage{caption}
\usepackage{subcaption}
\usepackage[font=small,skip=1pt]{caption}
\def\BibTeX{{\rm B\kern-.05em{\sc i\kern-.025em b}\kern-.08em
    T\kern-.1667em\lower.7ex\hbox{E}\kern-.125emX}}
\renewcommand\thesection{\arabic{section}} 

\renewcommand\thesubsectiondis{\thesection.\arabic{subsection}}

\renewcommand\thesubsubsectiondis{\thesubsectiondis.\arabic{subsubsection}}

\begin{document}

\title{\textbf{\Large Solsmith: Solidity Random Program Generator for Compiler Testing \\}}

\author{Lantian Li, Zhihao Liu, and Zhongxing Yu\\
	\normalsize Shandong University, Qingdao, China\\
	\normalsize lilantian@mail.sdu.edu.cn, zhihaoliu@mail.sdu.edu.cn, zhongxing.yu@sdu.edu.cn\
}


\maketitle
\begin{abstract}
Smart contracts are computer programs that run on blockchain platforms, with Solidity being the most widely used language for their development. As blockchain technology advances, smart contracts have become increasingly important across various fields. In order for smart contracts to operate correctly, the correctness of the compiler is particularly crucial. Although some research efforts have been devoted to testing Solidity compilers, they primarily focus on testing methods and do not address the core issue of generating test programs. To fill this gap, this paper designs and implements Solsmith, a test program generator specifically aimed at uncovering defects in Solidity compilers. It tests the compiler's correctness by generating valid and diverse Solidity programs. We have designed a series of unique program generation strategies tailored to Solidity, 
including enabling optimizations more frequently, 
avoiding undefined behavior, and mitigating behavioral differences caused by intermediate representations.
To validate the effectiveness of Solsmith, we assess the effectiveness of the test programs generated by Solsmith using the approach of differential testing. The preliminary results show that Solsmith can generate the expected test programs and uncover four confirmed defects in Solidity compilers, demonstrating the effectiveness and potential of Solsmith.
\end{abstract}
\IEEEoverridecommandlockouts
\vspace{1.5ex}

%
\IEEEpeerreviewmaketitle

\section{Introduction}
As programs that run on blockchain platforms, smart contracts enable transactions between parties without the need for a trusted third party. Owing to their features such as automated execution and operational transparency, smart contracts have been widely adopted in various fields, including cryptocurrency trading, supply chain management, and digital identity verification~\cite{ kannengiesser2021challenges,fsesolidity}. With the continued advancement of blockchain technology, smart contracts are expected to play an increasingly critical role in the future. Currently, a variety of programming languages can be used for smart contract development, among which Solidity is the most widely adopted and is regarded as the de facto standard for the Ethereum platform~\cite{antonopoulos2018mastering}. The syntax of Solidity was originally proposed by Gavin Wood in 2014 and supports modern programming features such as inheritance, libraries, and data structures~\cite{wood2014ethereum}. Smart contracts written in Solidity can be executed on the Ethereum Virtual Machine (EVM) and interact with other smart contracts and decentralized applications on the blockchain~\cite{alqarni2023use}.

As smart contracts grow in scale and complexity, the Solidity compiler plays an increasingly vital role in the contract development process~\cite{liu2023empirical}. Serving as a critical bridge between high-level source code and the low-level execution environment, the correctness and reliability of the compiler directly impact the security of smart contract execution. Although compiler developers are constantly working to improve stability and functionality, compilers, as complex software systems, are still susceptible to bugs~\cite{albert2022super}. Unlike ordinary software bugs, compiler errors may not only affect the compiler itself but also compromise a wide range of blockchain applications that depend on the generated code, potentially resulting in severe consequences. Worse still, developers typically assume that unexpected behaviors stem from issues in their own code. As a result, failure to recognize the compiler as the root cause may lead to considerable misdirected debugging efforts~\cite{chen2020survey}.

Automated testing~\cite{differentialtesting,multiple-fault,yuemse,yujss,YUgui} is one of the key methods for assessing compiler quality, and one of its core tasks is to generate a large number of diverse test programs to evaluate the behavior of the compiler~\cite{li2024boosting}. By automatically generating programs with various syntax structures, control flows, and data types, testers can comprehensively cover the different functions and boundary conditions of the compiler, thereby identifying potential compilation errors and inconsistent behaviors. The generation of these test programs is typically based on specific rules or random algorithms to simulate various scenarios encountered in real-world development, ensuring that the compiler produces the expected output when handling complex programs. Through such testing, the stability and reliability of the compiler are enhanced, thus reducing security risks and debugging costs caused by compiler bugs~\cite{chen2013taming, gu2023llm}.

Currently, research on Solidity compiler testing primarily focuses on the development of testing methods \cite{mitropoulos2023syntax,SQJCompilertesting,schumi2021spectest}, with limited studies on test program generation. Existing test program generators for other languages (such as Csmith~\cite{yang2011finding}) are effective at generating valid test programs, but these tools are typically tightly coupled with the specific language specifications, making them difficult to apply directly to Solidity. Therefore, designing and implementing a test program generator that can produce Solidity-compliant code is of significant research value. This paper fills this gap by proposing the Solidity random program generator, Solsmith. 

Through generating valid and diverse Solidity code, Solsmith is specifically designed to uncover defects in the Solidity compiler. We give how Solsmith is designed, implemented, and evaluated in this paper. First, we systematically discuss the design objectives and trade-offs for achieving consistency, diversity, and compliance in Solsmith. Then, in accordance with the design objectives and trade-offs, we give how Solsmith is implemented in detail. In particular, we have designed a series of unique program generation strategies tailored to Solidity. These strategies include more frequent activation of optimizations to make testing more efficient,
avoiding undefined behaviors, and preventing behavioral discrepancies
caused by intermediate representations to ensure consistency
of the programs across different environments. Finally, we conduct a preliminary evaluation of the effectiveness of Solsmith. We evaluate the effectiveness of the test programs generated by
Solsmith using the differential testing approach, and the
results show that Solsmith can generate the expected test
programs and reveal four confirmed defects in Solidity compilers.

This paper makes the following main contributions:
\begin{itemize}
\item We present the design objectives and trade-offs for the Solidity random program generator.
\item We design a series of unique code generation strategies that are particularly effective at exposing Solidity compiler bugs. 


\item We implement the Solidity random program generator Solsmith, which is open source and publicly available at \url{https://github.com/logicseek/Solsmith}.


\item We evaluate the effectiveness of the test programs generated by Solsmith using the mechanism of differential testing.
\end{itemize}


\section{Background}
\label{Background}
This section gives the necessary background on undefined behaviors and differential testing.
\subsection{Undefined Behaviors}
In the design of system programming languages, it is a challenging trade-off to determine the degree of freedom given to the compiler in generating efficient code for the target instruction set. On one hand, programmers desire consistent behavior of the program across all hardware platforms. On the other hand, programmers hope to achieve high performance by allowing the compiler to leverage specific attributes of the hardware platform's instruction set. One technique employed by language designers to strike this balance is to mark certain program structures as undefined behavior. Undefined behavior refers to actions in a program that involve code or data not adhering to the language specification, resulting in behavior for which the language standard does not explicitly define the outcome~\cite{liu2021kubo}. For example, in C language, an integer division by zero causes a hardware exception on the x86 architecture, while PowerPC simply ignores the operation. Instead of enforcing semantic uniformity across instruction sets, C defines division by zero as undefined behavior, allowing the C compiler to choose an appropriate implementation based on the target platform.

Compilers typically assume that programmers will not submit code with undefined behavior, and they optimize the code based on this assumption. However, programs containing undefined behavior may result in unexpected program behavior, as the compiler might remove certain code (e.g., eliminating access control checks) or rewrite code in ways that the programmer did not anticipate. If a program used to test a compiler contains undefined behavior, it is impossible to determine whether the compiler has a bug, even if the output does not match expectations.
Therefore, when generating test programs, it is important to avoid undefined behavior whenever possible.

\subsection{Differential Testing}
A compiler is inherently a complex piece of software, responsible for translating source code into executable machine code. Like any testing activity, compiler testing must address the oracle problem—how to determine whether a given test input results in unexpected behavior or errors.
Differential testing provides an effective solution to this issue~\cite{mckeeman1998differential}. The primary process of differential testing for compilers entails running generated test programs on multiple compilers. If the compilers yield differing results or crash, it indicates that a defect exists in one of the compilers.

Differential testing strategies can be categorized according to the relationships between the compilers being compared. The most commonly applied strategies include cross-compilation, cross-optimization, and cross-version strategies~\cite{evans2007differential}. The cross-compilation strategy, for instance, detects compiler errors by comparing the outputs produced by different compilers. Since different compilers should ideally generate the same output for identical input, discrepancies in output for the same input imply that one of the compilers is defective. The cross-optimization strategy, on the other hand, identifies potential errors by comparing the outputs of the same compiler under different optimization levels. Finally, the cross-version strategy detects errors by comparing results produced by different versions of the same compiler. Due to its efficiency and automation, differential testing not only facilitates the rapid identification of potential errors in compilers but also provides systematic error analysis, thereby offering substantial support for compiler enhancement and optimization~\cite{sharma2023rustsmith}.

\section{Design of Solidity Random Program Generator Solsmith}
\label{Design}
This section describes an overview of the design of the Solidity program random generator Solsmith.

\subsection{Design Overview}
\begin{figure}[htbp]
    \setlength{\belowcaptionskip}{-10pt} 
\centering
\includegraphics[scale=0.45]{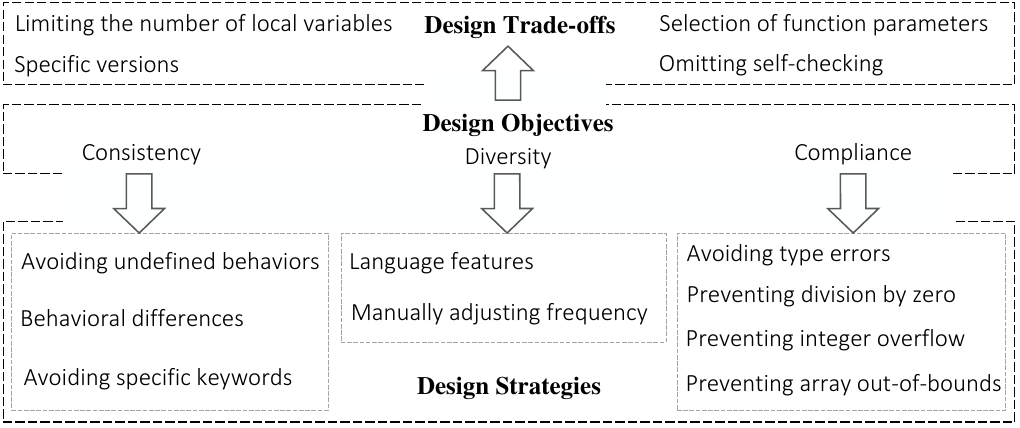}
\caption{Design Overview}
\label{fig:Overview}
\end{figure}

Figure \ref{fig:Overview} presents an overview of the design of Solsmith. The design of Solsmith is primarily focused on three key objectives: consistency, diversity, and compliance. First, consistency refers to the requirement that the generated test programs must have a single meaning across different environments. Test programs with multiple interpretations may lead to unpredictable effects on the test outcomes. The second objective is to maximize the diversity of the test programs. Diverse test programs can cover a wider range of code paths and execution scenarios, including various boundary cases and exceptional situations. The third objective is to ensure that each generated test program conforms to the Solidity language specification.

To achieve these three design objectives, we adopt corresponding design strategies. First, consistency is achieved by avoiding undefined behaviors, behavioral differences caused by intermediate representations (IR), and the use of specific keywords. Second, the diversity of the test programs is enhanced by combining a wide range of language features, along with manually adjusting the frequency of these features. Finally, we design a series of strategies to ensure the compliance of the test programs, including avoiding type errors, preventing division by zero, preventing integer overflow, and preventing array out-of-bounds errors.

On the other hand, we have made design trade-offs to address several key issues. These issues include: the inability to simultaneously satisfy all three design objectives in certain aspects of the implementation; syntactic differences across different versions of the Solidity language specification, which pose significant limitations for cross-version differential testing; the necessity of self-checking capabilities. The corresponding design trade-offs include limiting the number of local variables, adjusting the allowable nesting depth, choosing specific versions, and omitting self-checking capabilities.

\subsection{Consistency design}
\label{Consistency}
The primary goal of consistency design is to ensure that the generated test programs have a singular, well-defined meaning. A key factor contributing to the ambiguity of program meaning in Solidity is undefined behavior. Undefined behaviors in Solidity mainly include: length operations on storage-type arrays in inline assembly, function argument evaluation order, dangling references, and raw bytecode representations. The presence of these undefined behaviors can lead to unpredictable results, making the avoidance of such behaviors a crucial aspect of the consistency design discussed in this paper. Furthermore, another factor that contributes to the ambiguity of Solidity program meaning is the behavioral discrepancies caused by IR. The Solidity compiler supports two methods for generating bytecode: one directly from the Solidity source code and another through IR. These two methods introduce subtle semantic differences, which may lead to inconsistencies in program behavior. Behavioral discrepancies caused by IR in Solidity mainly include the repeated use of placeholders and inconsistent expression evaluation order. Thus, preventing behavioral discrepancies arising from IR is another important component of the consistency design.

To address the undefined behavior and behavioral discrepancies caused by IR, one approach is to generate programs in a structured manner to avoid such issues. However, in certain cases, avoiding these behaviors is not straightforward. In these situations, the Solsmith tool must ensure consistency in the test programs by preventing the introduction of specific keywords. Therefore, this paper primarily achieves its consistency goal through three methods: avoiding undefined behavior, eliminating discrepancies caused by IR, and ensuring consistency by preventing the introduction of specific keywords.

\subsection{Diversity design}
In order to enhance the coverage and depth of testing, the design of Solsmith necessitates the generation of test programs with diversity. A generator with rich diversity needs to support a wide range of language features and their combinations. The test programs generated by Solsmith include not only common syntax, such as \texttt{if/else} statements, function calls, \texttt{for} loops, \texttt{return}, \texttt{break}, and \texttt{continue}, but also the following Solidity-specific features.

Specifically, Solidity’s function modifiers are an important feature of the Solidity language. The test programs generated by Solsmith include several unique function modifiers, such as \texttt{payable}, \texttt{pure}, and \texttt{view}. The \texttt{payable} modifier allows a function to accept Ether and deposit it into the contract’s balance. The \texttt{pure} modifier indicates that the function does not access or modify the contract’s state, while the \texttt{view} modifier means the function can access but will not modify the contract’s state. These modifiers influence the behavior of the contract through specific functionalities. Additionally, the function selector in Solidity is another core concept. The function selector is used to uniquely identify a function signature, derived by hashing the function’s name and parameter types. In smart contracts, the function selector is primarily used to dynamically call functions in other contracts, allowing the contract to select and invoke specific functions during runtime as needed. This mechanism supports inter-contract communication and greatly enhances the flexibility and extensibility of contracts. Finally, Solidity has three primary data locations: \texttt{storage}, \texttt{memory}, and \texttt{calldata}. These are used to store different types of data. \texttt{Storage} is used to store state variables, \texttt{memory} is used to store local variables, and \texttt{calldata} is used to store function parameters, which cannot be modified. By supporting these features, Solsmith ensures that the test programs generated can cover all key functionalities of Solidity, thus improving the comprehensiveness and depth of the testing.

\subsection{Compliance design}
The normative design aims to ensure that the generated test programs strictly adhere to the specifications of the Solidity language. Next, we will provide a detailed explanation of our normative design.

\textbf{Preventing Type Errors.} If variables of different types are simply placed together in an arithmetic expression, a type error will occur at compile time. To enable more diverse arithmetic expressions and allow variables to participate in a wider variety of arithmetic operations, we perform explicit type conversions on variables with mismatched types, as illustrated in Figure~\ref{fig:type}. Specifically, when generating arithmetic expressions, we check whether the types of the selected variables are consistent with the type of the left-hand side value. If they are not consistent, we explicitly convert the type of the variable.

\begin{figure}[!htbp]
\setlength{\belowcaptionskip}{-15pt} 
\centering
\begin{lstlisting}[xleftmargin=2em, xrightmargin=2em]
function f() public {
    int256 a = 2;
    uint256 b = 5;
    b = b / uint256(a);
}
\end{lstlisting}
\caption{An Example of Explicit Type Conversion Insertion.}
\label{fig:type}
\end{figure}

\textbf{Preventing Division by Zero Errors.} To ensure correct implementation of division and modulo operations, it is essential to ensure that the divisor is non-zero, as a division by zero error will occur at compile time. Therefore, when selecting divisors for these two arithmetic operations, we introduce runtime checks, as shown in Figure~\ref{fig:Division}. This approach ensures that test programs generated do not violate Solidity standards.

\begin{figure}[!htbp]
\setlength{\belowcaptionskip}{-5pt} 
\centering
\begin{lstlisting}[xleftmargin=2em, xrightmargin=2em]
function f() public {
    uint256 a = 0;
    uint256 b = 0;
    b = b / (a == 0 ? 1 : b);
}

\end{lstlisting}
\caption{An Example of Preventing Division by Zero Errors.}
\label{fig:Division}
\end{figure}

\textbf{Preventing Array Index Out-of-Bounds Errors.} When an array index goes out of bounds, or when the \texttt{pop()} operation is used on an empty array, an array index out-of-bounds error will occur at compile time. Due to the presence of conditional and loop statements in the test programs, keeping track of array sizes becomes challenging. To address this, we adopt the strategy of inserting explicit checks for array lengths at the relevant statement positions when generating the test programs, as shown in Figure~\ref{fig:Array}. This ensures that the generated test programs will not result in array index out-of-bounds errors.
\begin{figure}[!htbp]
\setlength{\belowcaptionskip}{-12pt} 
\centering
\begin{lstlisting}[xleftmargin=2em, xrightmargin=2em]
contract C {
    uint256[] a = [1,2,3,4,5];
    function f() public {
        if(a.length>0)a.pop();
        if(a.length>5)a[5] ++;
    }
}
\end{lstlisting}
\caption{An Example of Preventing Array Index Out-of-Bounds Errors.}
\label{fig:Array}
\end{figure}

\textbf{Preventing Integer Overflow.} Prior to Solidity version 0.8.0, arithmetic operations would continue when an overflow occurs, and the result would default to the minimum or maximum value of the data type. However, starting from version 0.8.0, Solidity changed its handling of integer overflow, causing the arithmetic operation to revert. The loops and conditional statements introduced during Solsmith’s generation process make it difficult to control integer overflow. At the same time, we find that introducing loops and conditional statements significantly enhances the diversity of test programs. Therefore, we decide to retain these two syntax structures and use the \texttt{unchecked} keyword, as shown in Figure~\ref{fig:Integer}. This approach allows the compiler to revert to the overflow handling method used in versions prior to 0.8.0.
\begin{figure}[!htbp]
\setlength{\belowcaptionskip}{-5pt} 
\centering
\begin{lstlisting}[xleftmargin=2em, xrightmargin=2em]
function f() public {
    unchecked {
        uint256 a = 3;
        uint256 b = 5;
        a -= b;
    }
}
\end{lstlisting}
\caption{An Example of Preventing Integer Overflow.}
\label{fig:Integer}
\end{figure}

\subsection{Design trade-offs}
For the design of Solsmith, balancing factors such as objectives, constraints, and priorities is crucial. We next discuss the choices and trade-offs encountered during the design.

\textbf{Limiting the Number of Local Variables.} The fundamental reason for this decision is that, without limiting the number of local variables, it is easy to encounter the stack-too-deep error. This error stems from a limitation in the Ethereum Virtual Machine (EVM) regarding how variables are referenced in the stack. Although the EVM stack can hold more than 16 variables, attempting to reference variables in the 16th or higher stack slots results in this error. Therefore, it is necessary to limit the number of variables in the stack to no more than 16, leveraging block scope to address this issue. As shown in Figure~\ref{fig:Stack}, the original code contains 16 local variables. In the modified code, the scope of variable \texttt{p} is restricted within a new block. When the scope of this block is exceeded, variable \texttt{p} is removed from the stack, ensuring that variable \texttt{a} does not occupy a stack position higher than the 16th slot. Consequently, the modified code can execute correctly. However, note that this solution may impact the expressiveness of the test programs to some extent.

\begin{figure}[!t]
    \centering    
\begin{subfigure}[t]{0.22\textwidth}
        \centering
            \begin{lstlisting}[xleftmargin=1em, xrightmargin=0em]
function t() public {
uint256 a = 1;
uint256 b = 2;
...
uint256 o = 15;
uint256 p = 16;

uint256 number = a;
}
@ @
            \end{lstlisting}
        \caption{Original Code}
        \label{fig:test_program_3}
    \end{subfigure}
    \hfill
    \begin{subfigure}[t]{0.22\textwidth}
        \centering
            \begin{lstlisting}
function t() public {
uint256 a = 1;
uint256 b = 2;
...
uint256 o = 15;
{
    uint256 p = 16;
}
uint256 number = a;
}
            \end{lstlisting}
        \caption{Modified Code}
        \label{dce-5.8}
    \end{subfigure}  
    \hfill
    \caption{A Method to Avoid Stack-Too-Deep Errors.}
    \label{fig:Stack}
\end{figure}

\textbf{Adjusting the Allowable Nesting Depth.} Excessive function nesting can lead to issues such as stack overflow or performance degradation. Therefore, we can manually adjust the maximum allowable nesting depth. This adjustment allows us to maximize the expressiveness of the test programs while still adhering to the first design goal. When the maximum nesting depth is set to 1, even though function \texttt{g} can use either function \texttt{f} or function \texttt{g} as its parameters (as shown in line 8 of Figure~\ref{fig:Choice}), it is not allowed to choose function \texttt{f} or function \texttt{g} as its parameters, because the maximum nesting depth is 1 and function \texttt{g} is already a parameter of function \texttt{f}.
\begin{figure}[!htbp]
\setlength{\abovecaptionskip}{3pt}  
    \setlength{\belowcaptionskip}{-10pt} 
\centering
\begin{lstlisting}[xleftmargin=2em, xrightmargin=2em]
contract C {
    function f(uint256 a) public returns(uint256 res1){
        res1 = a;
    }
    function g(uint256 a) public returns(uint256 res2){
        res2 = a;
    }
    uint256 c = f(g(5));
}
\end{lstlisting}
\caption{Choice of Function Parameters.}
\label{fig:Choice}
\end{figure}

\textbf{Choosing Specific Versions.} The test programs generated by Solsmith are primarily designed for testing versions 0.8.0 and later. With the release of Solidity major version 0.8, earlier versions of the compiler will gradually cease to be maintained. Consequently, certain syntax from versions prior to 0.8.0 may not be supported. At the same time, we aim to test newer versions of the compiler, and thus Solsmith does not support older compiler versions.

\textbf{Omitting Self-checking Capabilities.} Before the test programs are executed, their output results cannot be predicted, so the test programs generated by Solsmith do not possess self-checking capabilities. Differential testing does not require test programs to have self-checking abilities, which allows us to utilize differential testing to validate the effectiveness of Solsmith. Admittedly, differential testing still has certain limitations. Some implementations within the compiler are quite similar or share common source code, which can result in identical error output under different configurations. If such a situation arises, we would not be able to detect the issue. This is due to an inherent limitation of differential testing itself, namely the lack of the ability to predict test outcomes.

\section{Implementation}
\label{Implementation}
This section presents the generation processes and strategies of Solsmith for test programs. 

\subsection{Generation Processes}
Solsmith first creates the environment for the entire test program, followed by the generation of the function framework. Each function is a top-level block that contains several code blocks, with each code block consisting of multiple nodes. Each node represents an operation from the list of statements. Once the number of nodes in a code block reaches the maximum limit, the current code block is considered complete. The next code block within the function is then created. When the number of code blocks in the function reaches the maximum limit, the function is considered complete. When the number of functions in the contract reaches the maximum limit, the entire contract generation is complete. The following will provide a detailed description of the generation processes.

\subsubsection{Environmental Construction}
The objective of this step is to generate the necessary components for smart contracts to ensure successful compilation and facilitate subsequent processes. The environment construction phase primarily consists of the following elements: the smart contract framework, event statement definitions, a set of state variables, dynamic arrays, and function modifiers. Each component will be elaborated in detail below.

First, the smart contract framework constitutes essential elements required for proper compilation, primarily including the Software Package Data Exchange (SPDX) license identifier and the smart contract name. SPDX provides a standardized method for accurately recording and communicating software license information in computer software and related documentation. Subsequently, events are employed to emit notifications and log messages within smart contracts, enabling communication between the contract and external systems. In Solsmith, events serve to record the contents of state variables, facilitating comparative analysis of their values during experimental verification.
The state variables generated by Solsmith are exclusively of \texttt{uint256} or \texttt{int256} types, each initialized with a randomly generated constant value. All state variable names and their corresponding types are stored in a dedicated state variable list. A subset of these state variables is marked with the \texttt{constant} modifier, and they are maintained in a separate list of constant state variables.
Furthermore, Solsmith generates arrays with randomized lengths, where all elements are strictly of \texttt{uint256} type and initialized with random constants. To enable enhanced array manipulation capabilities, all arrays in the test programs are implemented as dynamic arrays. The array names and initial lengths are recorded in an array variable list. These variables and arrays are subsequently combined with local function variables to form various expressions in the generated code.
Finally, modifiers are utilized to alter and constrain function behaviors, serving as preconditions for function execution validation, thereby reducing code redundancy. The modifiers generated by Solsmith incorporate expressions composed of state variables and arrays, along with at most one placeholder. These modifiers are randomly applied to functions, though some functions may not employ any modifiers.

\subsubsection{Function Framework Generation}
During the function framework generation phase, Solsmith employs randomized strategies to determine the attributes of generated function frameworks, including: the presence of return values, return types, function visibility, and modifier usage. After defining these attributes, Solsmith randomly selects statements for each node within the function from a predefined list containing the following operations: (i) Declaration of a new local variable with a randomly selected integer type (\texttt{uint256} or \texttt{int256}), initialized with a randomly generated constant value; (ii) Assignment of randomly generated arithmetic expressions to existing variables; (iii) Initialization of conditional statements; (iv) Creation of loop structures; (v) Embedding of inline assembly blocks; and (vi) Execution of array operations (e.g., \texttt{push()}).  

The phase concludes each function by invoking predefined event statements to log state variable values. These values are then compared across different compiler configurations to detect potential compiler defects. State variables are exclusively analyzed for comparison because only they persist on the blockchain after execution, whereas transient local variables leave no persistent impact post-runtime.

\subsubsection{Statement Filling}
The statements in Solsmith mainly consist of expressions, loop statements, conditional statements, and inline assembly statements. Expressions are composed of operands and operators and are used to compute a value. Variables used in expressions include state variables, arrays, and local variables. The types of expressions generated by Solsmith primarily include: arithmetic expressions, logical expressions, relational expressions, assignment expressions, function call expressions that return results, and index expressions used to access elements in arrays or mappings, among others.

The implementation of loop statements considers the limitation of nested loop depth and restricts the scope of the iteration counter to within the loop statement, thus avoiding compilation errors. With this design, loops of the same depth can share iteration counters with the same name, better controlling the number of iterations. There are two main forms of loop statements: (I) The initialization expression of the iteration counter appears at the beginning of the for loop, which means that the scope of the iteration counter is limited to within the loop body. The loop condition is typically set to check if the iteration counter is less than a random constant, and the iteration expression is used to increment the iteration counter; (II) The initialization expression of the iteration counter appears before the for statement, and the entire for loop (including the iteration counter) is placed within a code block. This method similarly restricts the scope of the iteration counter, making it valid only within that code block.

Conditional statements decide whether to execute a specific code block based on the truth of a condition. These conditions are described by logical expressions or relational expressions. A relational expression consists of two variables and a comparison operator, such as \texttt{a > b}. The comparison operators involved in the program generation process include \texttt{>}, \texttt{<}, \texttt{==}, and \texttt{!=}. A logical expression is composed of multiple relational expressions connected by logical operators, such as \texttt{(a \texttt{>} b) \texttt{\&\&} (c \texttt{>} d)}, which means that when both conditions are met at the same time, the corresponding code block is executed. The logical operators involved in the program generation process include \texttt{||}, \texttt{\&\&}, \texttt{!}, etc. Similar to loop statements, the generation of conditional statements is also limited by depth.

The language used for inline assembly in Solidity is called Yul. An inline assembly block consists of \texttt{assembly\{code block\}}, where the code in the curly braces is the code of the Yul language. Similar to ordinary Solidity code, each inline assembly block contains expressions, conditional statements, loop statements, and function-related statements that conform to the Yul specification.

\subsection{Generation Strategies}
To generate test programs that are more conducive to exposing defects in the Solidity compiler, relying solely on the code generation process presented earlier is insufficient. Therefore, we propose a series of unique test program generation strategies, which will be introduced in detail below.

\subsubsection{Frequent Optimization Activation}
To generate test programs more effectively, we adopted the approach proposed by John et al.~\cite{livinskii2020random}, which involves systematically distorting probability distributions to encourage certain optimizations to be triggered more frequently. This variation in probability distribution is applied to randomly selected sub-regions of the generated code. Due to the overlap of different generation strategies, Solsmith allows them to freely combine, thus introducing greater diversity into the generation process.

\textbf{Inline Assembly Optimization.} The Yul optimizer checks memory write operations in the Yul program, particularly those in outer Yul blocks. If a write operation is not read in subsequent code, the optimizer considers it unnecessary and removes it. This optimization applies to Yul code generated through the new via-IR pipeline, where the entire inline assembly block is optimized as a whole. In contrast, in the traditional code generation pipeline, if an inline assembly block does not reference surrounding Solidity variables, the Yul optimizer optimizes each inline assembly block independently. In this case, if a memory write operation is not read in subsequent instructions within the same inline assembly block, it will be deleted, even if other inline assembly blocks later access that memory. In the traditional code generation pipeline, inline assembly blocks that access Solidity variables do not trigger the Yul optimizer, thus significantly reducing errors caused by the optimizer mistakenly deleting necessary memory write operations. To trigger the Yul optimizer more frequently, we intentionally introduce local variables and state variables from Solidity within the inline assembly.

\textbf{Incorrect Removal of Storage Writes.} Optimization errors may occur if the contract includes the following steps: (1) A storage write; (2) A call to the \texttt{return()} or \texttt{stop()} function;
(3) Any continuing control flow path that overwrites the storage write from step (1) or triggers a rollback. When the initial storage write is potentially read between steps (1) and (3), it will not be removed. However, when the optimizer is enabled, determining whether a Solidity-level storage read is actually converted into a load instruction in the assembly is not always straightforward. For instance, the Load Resolver step may use the value previously written by the contract and directly replace \texttt{sload()} with the value to be read. In traditional code generation, these steps must occur within the same inline assembly block, and the \texttt{return()} or \texttt{stop()} must be calls to user-defined assembly functions. The Yul optimizer only runs when the inline assembly block does not reference Solidity variables, further reducing the occurrence of such errors. However, in the IR-based code generation process, the entire contract is first converted to Yul and optimized as a whole, which may cause steps (1) and (3) to appear in the Solidity code. As only the inline assembly block containing the call to \texttt{return()} or \texttt{stop()} triggers the optimizer, we intentionally introduce these two calls more frequently in the inline assembly block to increase the likelihood of such optimization errors.

\textbf{Keccak-256 Hash Optimization.} The \texttt{keccak256} function takes two parameters: an address \texttt{m} and a non-negative integer \texttt{p} that represents the length of the data to be read. The function computes the hash of the data from address \texttt{m} to \texttt{m + p - 1}. If the contents of the memory segment are known at compile time, the hash value computed by \texttt{keccak256} is treated as a constant. Suppose a program contains two \texttt{keccak256} calls whose hash values cannot be determined during compilation. If the optimizer determines that both calls access the same memory region, it treats the two hash operations as equivalent and generates the same result accordingly. Therefore, we tend to introduce the \texttt{keccak256} function more frequently in inline assembly during code generation, thereby triggering the optimizer more often and allowing it to identify more errors related to \texttt{keccak256}.

It is worth noting that the first approach (related to inline assembly optimization) requires more inline assembly code blocks to interact with Solidity variables, whereas the second approach (related to incorrect removal of storage writes) aims to minimize interaction with Solidity variables as much as possible. These two approaches are fundamentally opposed. To resolve this issue, we can artificially adjust the generation probabilities of different syntactic components, thereby increasing the likelihood of producing the desired code.

\subsubsection{Avoidance of Behavior Differences Caused by IR}
In the Solidity compiler, there are two distinct paths for generating EVM bytecode from the source code. The old code generator directly generates EVM bytecode from Solidity, while the new code generator first converts Solidity into Yul's IR and then generates the EVM bytecode. The new generator enhances the transparency and auditability of the code generation process and provides robust cross-function optimization channels. Since the Solidity compiler development team aims to discourage smart contract developers from relying on certain behaviors, there are subtle semantic differences between the old and new generation paths regarding these behaviors.

One of the differences arises from the repeated use of the placeholder “\texttt{\_;}”. In the old generator, each function's parameters and return values have fixed slots in the stack, and when the placeholder is used multiple times or within a loop, any changes to the parameters or return values will be visible during the next execution. In contrast, the new generator uses actual function implementation decorators, where function parameters are passed during each execution, and return values are reset to their default value (zero) during each execution. To avoid differences caused by the old and new code generators, we take measures to prevent such behaviors in the generated code. Specifically, a counter is defined and initialized to zero when generating decorators. Each time a placeholder is generated, the counter is incremented. When generating each statement, we check whether the counter is zero. If it is zero, it indicates that no placeholder has been generated before, and a placeholder can be generated; if it is not zero, it means a placeholder has already been generated, and a new random statement is selected until no placeholder exists.

Another behavioral difference arises from inconsistencies in expression evaluation order. The old generator does not specify a defined order for evaluating operands or sub-expressions, whereas the new generator generally follows a left-to-right evaluation strategy, but this order is not strictly guaranteed. As a result, the same expression may produce different outcomes depending on the evaluation order. To avoid this issue, we use a hash table to track the variable states when generating arithmetic expressions. At each selected variable position, we randomly insert increment or decrement operations, and determine the operation type based on the hash table values. If a variable was previously used alone, it can continue to be used alone; however, if the variable has participated in an increment or decrement operation, a randomly generated constant will replace it if it has been selected again.

\subsubsection{Avoidance of Undefined Behaviors}
Undefined behavior refers to behavior in a program that is not explicitly defined under certain conditions, usually to consider factors such as performance or portability. This means that any action the compiler takes in response to encountering undefined behavior is considered correct. To prevent such situations while testing the Solidity compiler, we conducted a search for the term ``undefined behavior'' in the official Solidity documentation and its GitHub repository, filtering the results. We identified four types of undefined behavior and took measures to avoid them in the generated test programs.

The first undefined behavior involves performing length operations on storage-type arrays within inline assembly. Storage slots are used to store contract states, and the Solidity compiler typically keeps unused storage slots empty. When using \texttt{pop()} to remove elements from an array, the compiler explicitly writes zero to the storage slot of the popped value. However, when directly shortening the array length within inline assembly, the remaining storage operations are uncontrolled. To avoid this, we refrain from manipulating arrays within inline assembly.

The second undefined behavior concerns the evaluation order of function parameters. As shown in Figure~\ref{fig:ParameterOrder}, if the parameter evaluation order is from left to right, the initial value of variable \texttt{x} is not 1337, so the condition evaluates to true, triggering the \texttt{set} function to set \texttt{x} to 1337, and the final return value of the function \texttt{f} becomes 1337. If the evaluation order is from right to left, the \texttt{set} function is executed first, setting \texttt{x} to 1337. This makes the condition false, causing the contract's execution to stop and the state to roll back. Therefore, we avoid relying on the parameter evaluation order in function definitions.
\begin{figure}[!htbp]
\setlength{\belowcaptionskip}{-10pt} 

\centering
\begin{lstlisting}[xleftmargin=1em, xrightmargin=0em]
uint x;
function set() public returns(string memory s) {
    x = 1337;
}
function f() public returns (uint) {
    require(x != 1337 , setValue());
    return  x;
}
\end{lstlisting}
\caption{An Example of Function Parameter Evaluation Order.}
\label{fig:ParameterOrder}
\end{figure}

The third undefined behavior refers to dangling references. In Solidity, the storage layout for dynamic arrays and strings includes both large layout and short layout. In short layout, both the data and its length are stored within the same storage slot, making it suitable for relatively small data. In contrast, the large layout separates the storage of the data and its length. When a string or dynamic array is stored using the short layout, and its length increases beyond a certain threshold, the storage format transitions to the large layout. As a result, a reference previously created based on the short layout may become invalid, potentially pointing to incorrect or unintended data. This leads to undefined behavior due to a dangling reference. To avoid this, we only expand dynamic arrays and strings by one element during assignment and avoid accessing dynamic arrays or strings within the same statement.

The fourth undefined behavior pertains to verbatim bytecode. The built-in Solidity function \texttt{verbatim} allows developers to generate arbitrary opcodes that are unknown to the Solidity compiler and are not subject to optimization. Improper use of \texttt{verbatim} can affect the correct stack layout, resulting in undefined behavior. Therefore, we avoid using \texttt{verbatim}-related functions in Solsmith to ensure that bytecode operations adhere to the expected stack layout and control flow.

\section{Experimental Results and Analysis}
\label{Experimental}
This section presents the evaluation results of Solsmith. 

\subsection{Experimental Process}
Solc is the current standard, official Solidity compiler and it offers a wide range of configuration options, including settings related to the optimizer and bytecode generation~\cite{soliditylang}. We leverage the optimization-related compiler options to implement a cross-optimization strategy. Specifically, we run the test programs generated by Solsmith under different compiler configurations and compare the results. If a test program produces different results under different compiler configurations or causes the compiler to crash, it indicates that there is a potential issue with certain compiler configurations. To facilitate switching between different versions and configurations of the Solidity compiler, we utilize the Hardhat framework~\cite{hardhat}.


Our experimental configuration includes the following three setups: the first setup is the standard configuration, where no optimizations are enabled; the second setup enables all optimizations and targets optimization for initial deployment costs, with the \texttt{runs} attribute set to its minimum value of 1; the third setup enables all optimizations and targets optimization for high-frequency calls, with the \texttt{runs} attribute set to its maximum value of $2^{32} - 1$.

\subsection{Experimental Results}
In the experiment, we generated a total of 100,000 test programs. The program with the largest number of lines contains 808 lines and the one with the smallest number of lines has 70 lines, with an average of 169.1 lines per program. The generated test programs were used to evaluate the compiler (i.e.,solc), revealing four distinct, confirmed defects in version 0.8.0. The four defects were distributed across three different compiler configurations, and we next provide a detailed discussion of them.

\subsubsection{The FullInliner Defect}
This defect is triggered by the program shown in Figure~\ref{fig:FullInliner}. When the compiler is configured with the standard settings, the program does not produce any output. However, enabling all optimizations results in the error message: ``Transaction reverted without a reason string.'' The issue originates from the optimizer component FullInliner, which is responsible for replacing certain function calls with the function body. In specific scenarios, such as when function parameters are constants, this replacement can positively affect subsequent optimizations. The defect is triggered when the following four conditions are met: (I) The Yul optimizer is enabled; (II) A custom sequence of optimizer steps is used; (III) The sequence includes a FullInliner step; (IV) The non-expression-splitted code reaches the FullInliner step.
\begin{figure}[!htbp]
\centering
\begin{lstlisting}[xleftmargin=2em, xrightmargin=2em]
function f(x,y)  -> r {
    let c := mul(x,4)
    r := add(c,y)
}
function ret() -> r { return(0,0)}
function rev() -> r { revert(0,0)}
let a := f(ret(),rev())
\end{lstlisting}
\caption{A Program that Triggers the FullInliner Defect.}
\label{fig:FullInliner}
\end{figure}

Expression splitting in Condition (IV) refers to decomposing complex expressions into simpler components to facilitate later code processing. The original code, without this decomposition, is referred to as non-expression-splitted. The cause of the defect lies in FullInliner’s failure to preserve the evaluation order of function parameters during inlining. When the function parameters themselves are function calls with side effects, changes in evaluation order can alter the contract's execution outcome. By default, this step follows ExpressionSplitter, which converts the code into expression-split form, ensuring that parameters are simple identifiers and avoiding evaluation order issues. However, as Solidity evolved, a mechanism was introduced to allow users to provide custom optimization sequences, leading to the triggering of this defect under certain compiler configurations.

\subsubsection{The Selector Defect}
This defect involves function calls between different smart contracts. When the \texttt{viaIR} option is enabled in the compiler configuration, running the code shown in Figure~\ref{fig:Selector} results in behavior different from the standard configuration. Enabling \texttt{viaIR} uses an IR-based code generator, which assigns the value 42 to \texttt{x} and creates an instance of smart contract \texttt{D}. If \texttt{viaIR} is disabled, the traditional code generator is used, which results in \texttt{x} being 0 and no instance of smart contract \texttt{D} being created. Function \texttt{f} calls \texttt{h}, which assigns the value 42 to \texttt{x} and creates an instance of \texttt{D}, and then uses this instance to call function \texttt{g} in contract \texttt{D}. However, due to the defect, the expected behavior cannot be achieved.
\begin{figure}[!htbp]
\setlength{\belowcaptionskip}{-10pt} 
\centering
\begin{lstlisting}[xleftmargin=2em, xrightmargin=2em]
contract D {
    function g() external {}
}
contract C {
    uint256 x;
    function f() public {h().g.selector;}
    function h() public returns (D) {
        x = 42;
        return new D();
    }
}
\end{lstlisting}
\caption{A Program that Triggers the Selector Defect.}
\label{fig:Selector}
\end{figure}

When the result of a function can be determined at compile time, the traditional code generator incorrectly assumes that the expression will not have additional side effects. This leads to the function selector being calculated without evaluating the expression, generating more efficient code. In contrast, the new IR-based code generator prioritizes generating correct code, deferring efficiency improvements to the Yul optimizer. This means that the IR-based generator does not make such assumptions, and the unoptimized code always accesses and evaluates the entire expression. The defect also affects expressions whose values cannot be determined at compile time. Internally, the compiler associates function types with their definitions and passes the relevant definitions to the result type of the expression. If the expression involves multiple functions, the result type may point to the wrong function.

\subsubsection{The Keccak256 Defect}
This defect occurs when calculating the \texttt{keccak256} hash value. To trigger this defect, the \texttt{cse} option in the compiler configuration must be enabled. Additionally, the test case must meet the following conditions: (I) there are two \texttt{keccak256} functions in the code; (II) the first parameter of the two \texttt{keccak256} functions is the same; (III) the second parameter of the two functions is rounded up to the nearest multiple of 32 and is the same; (IV) there should be no instructions that break the control flow between the two \texttt{keccak256} functions, such as function calls, if statements, etc. The code example shown in Figure~\ref{fig:keccak256} triggers this defect.
\begin{figure}[!htbp]
\centering
\begin{lstlisting}[xleftmargin=2em, xrightmargin=2em]
contract C {
    function f() public returns (uint a,uint b) {
        assembly {
            mstore(0,0)
            a := keccak256(0,32)
            b := keccak256(0,23)
        }
    }
}
\end{lstlisting}
\caption{A Program that Triggers the keccak256 Defect.}
\label{fig:keccak256}
\end{figure}

The root cause of the issue lies in the optimizer erroneously reusing a previously computed hash value. The Solidity bytecode optimizer has a step where, if the memory content to be hashed is known during a \texttt{keccak256} function call, it computes the hash at compile time. However, this step contains a flaw: the optimizer incorrectly considers \texttt{keccak256} hashes with the same memory content but different sizes to be equal. Furthermore, it is important to note that in Solidity, besides directly using the \texttt{keccak256} function, the compiler may also generate code that calls \texttt{keccak256}. For example, \texttt{keccak256} calls generated by the compiler may be triggered when using indexed events, using strings or bytes as mapping keys, or accessing mappings and arrays.

\subsubsection{The Unchecked Defect}
This defect is related to the \texttt{unchecked} keyword. In Solidity version 0.8.0, overflow checks are enabled by default for integer operations, and an exception is thrown to interrupt the execution if an overflow is detected. The \texttt{unchecked} keyword disables these checks, thus improving computational speed. The example program shown in Figure~\ref{fig:Unchecked} triggers this defect. Under the standard configuration, the state variable \texttt{a} in line 5 is of type \texttt{uint256} and is placed inside the \texttt{unchecked} block, so it is expected to be set to the maximum value of \texttt{uint256}. After executing \texttt{a++}, \texttt{a} becomes 0. However, when \texttt{viaIR} is enabled, the compiler throws an overflow error.
\begin{figure}[!htbp]
\centering
\begin{lstlisting}[xleftmargin=2em, xrightmargin=2em]
contract C {
    uint256 a = 42;
    function f() public {
        unchecked {
            a = a - 1 - a;
            a ++;
        }
    }
}
\end{lstlisting}
\caption{A Program that Triggers the Unchecked Defect.}
\label{fig:Unchecked}
\end{figure}

In fact, if \texttt{a} is assigned the minimum value of \texttt{uint256} and then \texttt{a--} is executed, the same error is thrown. Replacing \texttt{a++} with \texttt{a+=1} or \texttt{a=a+1} allows the code to compile successfully. Therefore, it can be concluded that the code generated via IR cannot properly handle the increment operation for the maximum value and the decrement operation for the minimum value of unsigned integers.
\section{Related Work}
\label{Related}
\subsection{Compiler Testing}
Compiler testing has become a prominent area of research in recent years~\cite{zang2022compiler,compileroptimization}. 
Gu proposed a code generation method based on large language models (LLM) for compiler testing, aiming to maximize the quality and quantity of generated code~\cite{gu2023llm}. 
Donaldson et al. proposed a transformation-based compiler testing method, which reduces defect-inducing transformation sequences to minimal subsequences~\cite{donaldson2021test}.
 Chaliasos et al. introduced a testing framework for validating static typing procedures in compilers~\cite{chaliasos2022finding}. 
Liu et al. proposed a fuzz testing approach to detect defects in deep learning compilers~\cite{liu2023nnsmith}. 
The Solidity language is relatively new, and thus there has been limited research on testing Solidity compilers. Mitropoulos et al. proposed FUZZOL, the first syntax-aware mutation fuzzer designed to test the security and reliability of the Solidity compiler~\cite{mitropoulos2023syntax}. 
Schumi et al. introduced a novel specification-based testing method for the Solidity compiler, SpecTest, with a new semantic coverage criterion~\cite{schumi2021spectest}. 

\subsection{Test Program Generation}
The test program is vital for compiler testing~\cite{li2024boosting}. For statically-typed languages, constructing a valid program that strictly adheres to syntax rules, type constraints, and other specifications is not an easy task~\cite{chen2020survey}. 
To build highly diverse test suites, Chen et al. used the distance between test programs as a measure of diversity~\cite{chen2013taming}. Berry et al. proposed a method for compiler test program generation based on the frequency of language features used by real programmers~\cite{berry1983new}.
Mandl et al. proposed a method to avoid generating duplicate elements in an Ada program~\cite{mandl1985orthogonal}. 
Existing research on test program generation primarily focuses on mainstream languages,
and there is a lack of research on techniques for the Solidity language.

\section{Conclusion}
\label{Conclusion}
This paper designs and implements a test program generation tool, Solsmith, for testing the Solidity compiler. We ensure the consistency of the test programs by adopting measures such as avoiding undefined behavior and mitigating behavioral differences caused by IRs. The tool achieves diversity in the test programs by supporting a wide range of language features and their combinations, as well as allowing manual adjustments to the frequency of feature occurrences. Meanwhile, we ensure the compliance of the generated programs by preventing errors such as type mismatches and integer overflows. 
We evaluate the effectiveness of the generated test programs based on the mechanism of differential testing, and the results are promising. Future work will focus on exploring more diverse variable types, more complex contract inheritance relationships, and testing under a wider range of optimization combinations to cover more of the testable content in the Solidity compiler.





\balance 
\bibliographystyle{ieeetr}
\bibliography{references}

\end{document}